\documentclass[twocolumn,showpacs,preprintnumbers,prd,nofootinbib]{revtex4-1}
\usepackage{latexsym}
\usepackage{graphicx}
\usepackage{subfig}
\usepackage[spanish]{babel}

\begin{document}
\title{Materia oscura escalar compleja (parte I): la versi\'on hidrodin\'amica}
\author{Mario A. Rodr\'iguez-Meza$^1$, Alberto Hern\'andez-Almada$^2$ y Tonatiuh Matos$^2$}
\affiliation{
$^1$Departamento de F\'{\i}sica, Instituto Nacional de
Investigaciones Nucleares, Apdo. Postal 18-1027, M\'{e}xico D.F. 11801,
M\'{e}xico.
e-mail: marioalberto.rodriguez@inin.gob.mx \\
$^2$Departamento de F\'\i sica, Centro de Investigaci\'on 
y de Estudios Avanzados del IPN, 
A.P. 14-740, 07000 Mexico City, Mexico. 
}

\begin{abstract}
En este trabajo
utilizamos las ecuaciones hidrodin\'amicas de Euler
para modelar halos de materia oscura escalar compleja en el Universo, las cuales adquieren la forma de un sistema Schr\"odinger-Poisson en el
l\'imite Newtoniano. Mediante la transformaci\'on de Madelung, dicho sistema adopta la
forma de la din\'amica de un  fluido, en donde interviene un potencial de auto-interacci\'on
y un potencial tipo cu\'antico que depende no linealmente de la densidad del 
fluido. En este marco te\'orico se hace un an\'alisis de inestabilidad de Jeans, el cual
sirve para determinar el tama\~no m\'inimo para que el sistema sea estable, es decir,
para que 
perturbaciones del campo escalar formen estructuras.
Tambi\'en mostramos que este sistema hidrodin\'amico de campo escalar tiene vorticidad.

\medskip
\noindent
\textit{Descriptores}: 
Campos escalares; materia oscura; modelo de fluido; inestabilidad de Jeans; vorticidad hidrodin\'amica

\bigskip
\noindent
In this work 
we use the Euler hydrodynamic equations of fluids to study a model of galactic halos  minimally coupled to a complex
scalar field, which in the
Newtonian limit they become the Schr\"odinger-Poisson system. Applying a Madelung transformation,
this system of equations takes the form of hydrodynamics equations, 
where there are a self-interacting
potential and a kind of quantum potential that depends non-linearly  on the density
of the fluid. In this theoretical framework we analyze the Jeans' instability, which is
useful for finding the scale length 
of perturbations of the scalar field that will form structures.
In other words, perturbations
of the scalar field with lengths less than this threshold length,
can not lead to the formation of galactic structures. We also show that this scalar field hydrodynamic system has vorticity.

\end{abstract}

\date{28 de julio de 2011}
%
% PACS Numbers (ver 2008)
% 95.30.Lz			; Hydrodynamics
% 95.30.Sf			; Relativity and gravitation
% 95.35.+d			;  Dark matter (stellar, interstellar, galactic, and cosmological)
% 98.62.Gq			; Galactic halos
% 98.80.Jk			; Mathematical and relativistic aspects of cosmology
% 04.25.Nx			; Post-Newtonian approximation; perturbation theory; related
%						approximations
%
\preprint{}
\maketitle

%%%%%%%%%%%%%%%%%%%%%%%%%%%%%%%%%%%%%%%%%%%%
%%%%%%%%%% SECTION INTRODUCCION%%%%%%%%%%%%%%%%%%%%%%%%
\section{Introducci\'on}

En las \'ultimas d\'ecadas se ha vivido una revoluci\'on en el conocimiento humano debido a la 
aparici\'on de dos nuevos problemas fundamentales en la cosmolog\'ia moderna, ambos
relacionados con la formaci\'on de estructura en el Universo. El primero es una componente
 necesaria para la 
formaci\'on de estructura, es decir, para la formaci\'on de galaxias, c\'umulos de galaxias, etc. Este
 primer ingrediente es al menos el $23\%$ de la materia del cosmos y es generalmente llamada
  \textit{materia oscura}\footnote{Se dice oscura porque s\'olo interacciona gravitacionalmente con 
  la materia ordinaria referida en la literatura como la materia bari\'onica.}. La segunda componente 
  es 
una componente 
  que acelera el Universo constantemente. Esta \'ultima se le conoce como \textit{energ\'ia oscura} y 
  corresponde al $73\%$ de la materia del Universo y la tercera, que tan s\'olo representa el $4\%$, corresponde a la materia
  bari\'onica contenida en el modelo est\'andar \cite{Hinshaw:2009, Dunkley:2009, Komatsu/others:2010}. Desde la aparici\'on de estos 
  dos primeros ingredientes, una gran cantidad de cient\'ificos se han concentrado en analizar decenas de 
  hip\'otesis sobre su posible naturaleza y origen, sin embargo, los resultados no son del todo 
  satisfactorios y el misterio hasta ahora contin\'ua.

Una diferencia sustancial entre la materia oscura y la energ\'ia oscura es que la primera es
 gravitacionalmente atractiva mientras que la segunda es repulsiva, lo cual es verdaderamente 
 extra\~no. 
 En el marco del modelo est\'andar, no se conoce ninguna part\'icula elemental 
 con propiedades tales que produzcan efectos gravitacionales repulsivos.
 Por lo tanto las 
 preguntas m\'as importantes a resolver son \textquestiondown Qu\'e son estas 
 componentes oscuras que dominan en el Universo? \textquestiondown Cu\'al es la 
 naturaleza de estas componentes? Existen varias hip\'otesis que quieren dar una respuesta a 
 estas interrogantes, la pro\-pues\-ta m\'as popular es que la materia oscura 
 es una clase de part\'icula supersim\'etrica llamada WIMP, por su siglas en ingl\'es 
 ``Weakly Interacting Massive Particle''. Para la energ\'ia oscura, el candidato m\'as popular  es 
 sin duda la constante cosmol\'ogica, denotada por $\Lambda$. Sin embargo, ninguno de estos candidatos est\'an exentos de problemas. 
 Por ejemplo los WIMP's siempre se comportan como polvo en la galaxia, 
  por lo que no hay un mecanismo real que impida que se formen halos de galaxias de cualquier
   tama\~no y no hay nada que impida su colapso gravitacional. 
  Es decir, la predicci\'on del modelo $\Lambda$CDM en la formaci\'on de estructura es que los 
  halos de materia oscura crecen jer\'arquicamente,  por acreci\'on y fusi\'on de sistemas m\'as 
  peque\~nos a cada vez m\'as grandes. Una consecuencia de esto es que la V\'\i a L\'actea actualmente
  deber\'\i a estar rodeada por miles de peque\~nos subhalos, en aparente contradicci\'on
  con los sat\'elites luminosos peque\~nos  que han sido detectados alrededor de la 
  V\'\i a L\'actea hasta ahora.
   A este problema se le conoce 
   como \textit{problema de los sat\'elites faltantes}
\cite{Font:2011,Klypin:1999, Moore:1999}.
Por otro lado la constante cosmol\'ogica es simplemente un par\'ametro que la teor\'ia puede
permitir pero que no est\'a determinado. Mide la densidad de energ\'ia del espacio vac\'{i}o.
Pero para lograr obtener una expansi\'on acelerada del Universo en nuestros d\'ias,
se requiere que la constante cosmol\'ogica sea del orden $10^{-47}$ GeV$^4$. Si calculamos la
densidad de energ\'ia del vac\'{i}o, calculada como la suma de energ\'ias de punto cero de un
campo cu\'antico de cierta masa, obtenemos que la raz\'on de la densidad del vac\'{i}o que la teor\'ia
est\'andar de part\'iculas da a la densidad observada es del orden $10^{121}$. Si usamos
supersimetr\'ia se puede reducir esta raz\'on en 60 \'ordenes de magnitud, lo que sigue siendo
insuficiente\cite{Sapone2010}.

\subsection{Materia oscura fr\'ia}

Los WIMP's son el prototipo para  el modelo de materia oscura fr\'ia 
(CDM por sus siglas en ingl\'es) \cite{Blumental/etal:1984}. Sus predicciones se encuentran 
en concordancia con las observaciones a grandes escalas
 ($\sim$ Mpc)\footnote{1 p\'arsec (pc) $\approx$3.26 a\~nos luz}, 
 y logra explicar diversas propiedades de las galaxias y c\'umulos de galaxias.
 Sin embargo, 
 a pesar de los grandes \'exitos de CDM, \'este presenta discrepancia a escalas 
 menores ($\sim$ kpc). Una de ellas consiste en que los perfiles de densidad predichos 
 por CDM son demasiado densos en el centro de las galaxias.
En contraste,
observaciones muy precisas muestran que las galaxias, sobre todo las enanas o 
las de muy bajo brillo, tienen densidades m\'as bien con un perfil plano o casi plano.
A este problema se le conoce en la literatura como
  el \textit{problema de los halos picudos}
\cite{Klypin:1999, Moore:1999}. 

El problema de la subestructura y el de los halos picudos 
se consideran
fundamentales 
para entender la formaci\'on de estructura en el Universo \cite{Klypin:1999, Moore:1999}. Incluso 
si el modelo CDM tiene \'exito, resolver estos problemas ser\'a de crucial importancia para 
este paradigma. Adem\'as, recientemente han surgido otras observaciones que tambi\'en 
pueden ser un problema para el paradigma de la materia oscura fr\'ia. \'Este afirma que la estructura se form\'o 
de una ma\-ne\-ra jer\'arquica, es decir que la estructura surgi\'o de peque\~nas semillas de 
excesos de densidad de materia oscura, se cree que de masas como la de la Tierra y del 
tama\~no del sistema solar. 
Estas semillas se 
fusionan
por su propia gravitaci\'on, pero al ser 
el Universo primitivo tan denso, las semillas chocan con otras a su alrededor y muchas 
veces se 
fusionan
con otras para formar halos m\'as grandes. Estos halos mayores contin\'uan 
chocando con otros hasta formar los halos que ahora conocemos. En este paradigma, 
las galaxias m\'as grandes se forman por 
la fusi\'on
gravitacional de otras m\'as 
peque\~nas. Sin embargo, han surgido varias observaciones que podr\'ian contravenir 
estas hip\'otesis. Las primeras observaciones consisten en el an\'alisis de cientos de 
galaxias \cite{Disney/others:2008}, en donde 
se encuentra
que seis de las caracter\'isticas de las galaxias, como son brillo superficial, masa, 
tama\~no del halo, etc. est\'an correlacionadas, es decir, si se conoce una de \'estas 
se pueden deducir las otras propiedades restantes. Esta observaci\'on no puede ser 
explicada con el modelo jer\'arquico, pues no hay manera de obtener galaxias con 
caracter\'isticas tan semejantes como las que se observan, despu\'es de una serie de 
choques y fusiones de halos de forma pr\'acticamente aleatoria. Es mucha casualidad 
que dos galaxias del mismo tama\~no sean tan semejantes, despu\'es de que ambas 
tuvieron un historial de choques y fusiones que podr\'ian ser completamente distinto. 
Con la hip\'otesis del modelo jer\'arquico estas semejanzas son muy dif\'iciles de explicar. 
M\'as aun, recientemente se encontr\'o que la fuerza de gravedad de la materia oscura de 
todas las galaxias a un radio determinado es la misma para todas las 
galaxias \cite{Gentile/others:2009}, una semejanza universal dif\'icil de explicar con el 
modelo jer\'arquico.

La segunda observaci\'on que parece contravenir el modelo jer\'arquico 
consiste en que al examinar galaxias con los nuevos telescopios 
gigantes de reciente construcci\'on, seg\'un el modelo jer\'arquico, las galaxias m\'as 
grandes tardaron mayor tiempo en formarse. Por otro lado, simulaciones num\'ericas
\cite{Collins/others:2009} 
 demuestran con gran detalle el tama\~no de las galaxias que se formaron en cada 
 etapa del Universo. Al hacer una comparaci\'on de estas simulaciones con algunas 
 galaxias formadas en tiempos correspondientes a $z=1\mbox{.}3$, 
 en \cite{Collins/others:2009} encuentran una gran discrepancia entre lo observado 
 y lo predicho. Esta observaci\'on apunta al hecho de que las galaxias se formaron 
 m\'as temprano a lo predicho por la hip\'otesis de la materia oscura fr\'ia.

La tercera observaci\'on es sobre la dispersi\'on de velocidades 
de galaxias a corrimientos al rojo 
muy grande, por ejemplo, $z=2\mbox{.}186$, 
es decir galaxias muy viejas. Muchas de estas galaxias
son muy compactas y con masas estelares similares a las galaxias el\'\i pticas de hoy d\'\i a pero con tama\~nos m\'as peque\~nos.
En la referencia
\cite{Dokkum/others:2009} 
se 
analiz\'o una de estas galaxias y se encontr\'o que la dispersi\'on de velocidades es muy alta,
$500^{+165}_{-95}$ km s$^{-1}$, lo que es consistente con lo compacto y masivo de la
galaxia. La relaci\'on inferida entre el tama\~no, $r_e$, la masa din\'amica, $M_{din}$, y la 
dispersi\'on de velocidades, $\sigma_{v}$, es $\sigma_{v}^2 \propto M_{din} / r_e$.
Es decir,
la dispersi\'on de las velocidades del gas y de las estrellas de estas galaxias 
indican que las galaxias el\'\i pticas 
ya ten\'ian la masa que actualmente tienen, pero ya
estaban m\'as concentradas y  peque\~nas. Las galaxias parecen
 haberse formado muy temprano, y estas observaciones sobre la dispersi\'on de 
 velocidades indican que eran m\'as compactas. Si estas observaciones se confirman, 
 indicar\'ian una fuerte contradicci\'on con el modelo de materia oscura fr\'ia. De cualquier 
 forma, si se descubre alguna part\'icula supersim\'etrica que pueda servir de sost\'en a la 
 materia oscura fr\'ia, este modelo deber\'a resolver todas estas anomal\'ias existentes con 
 las observaciones. 

\subsection{Materia oscura escalar}

Para darle vuelta a estas fallas en el modelo de materia oscura fr\'ia han surgido algunas
alternativas, 
entre las que se encuentra la hip\'otesis de materia oscura 
escalar \cite{Guzman/Matos:2000, Guzman/Urena:2003, Matos:2010}. Esta teor\'ia tiene 
algunas caracter\'isticas muy interesantes que podr\'ian resolver algunas anomal\'ias de 
la materia oscura fr\'ia. A pesar de que la hip\'otesis de la materia oscura escalar es casi 
id\'entica al modelo de materia oscura fr\'ia a escalas cosmol\'ogicas, tiene fuertes 
diferencias a escalas gal\'acticas. La idea de este modelo es la siguiente. En un principio 
exist\'ia un campo escalar con una masa muy ligera del orden 
de $m_\phi \sim 10^{-22}e\mbox{V}$, esparcido homog\'eneamente en el Universo. 
Despu\'es de la inflaci\'on del Universo, este campo escalar se condens\'o a di\-fe\-ren\-tes 
escalas y de alguna manera inici\'o un colapso gravitacional. Aqu\'i surge la primera 
diferencia sustancial con el modelo de materia oscura fr\'ia. Debido a que las semillas 
de los halos de las galaxias se formaron por condensaci\'on, estos halos se esperan de 
todos los tama\~nos. Las fluctuaciones primordiales inician el proceso de colapso 
gravitacional de estos halos en un tiempo muy temprano y por eso vamos a ver halos 
bien formados en tiempos anteriores a los predichos por la materia oscura fr\'ia. Este 
esquema es s\'olo hipot\'etico, uno de los objetivos de este proyecto es ver si este es el 
caso. Se espera entonces que el campo escalar evolucione. 
Ya sabemos que en 
la \'epoca de fluctuaciones lineales este campo se comporta exactamente como materia 
oscura fr\'ia \cite{Matos/Urena-Lopez:2001}, pero el colapsarse y formar halos de galaxias 
es un proceso muy diferente al paradigma en boga. Este estudio ya se ha elaborado para 
un halo esf\'erico \cite{Guzman/Urena:2003, Guzman/Urena:2004, Guzman/Urena:2006}, 
la idea es llevar a cabo un estudio num\'erico completo, que nos pueda dar una gu\'ia 
 para ver si este modelo es correcto o debemos modificarlo.

En este trabajo 
reformulamos el
modelo de materia oscura escalar para estudiar halos de materia 
oscura de galaxias. 
Para este caso, se sabe que la din\'amica tanto del campo escalar 
como de la geometr\'ia del espacio-tiempo se describen bien en el l\'imite newtoniano y 
al sistema resultante se le conoce 
como \textit{Schr\"odinger-Poisson} \cite{Guzman/Urena:2003}. 
Algunos estudios de este sistema los podemos encontrar 
en \cite{Guzman/Urena:2003, Bernal:2006it, Guzman/Urena:2004}. 
Sin embargo, uno de los inconvenientes de resolver la ecuaci\'on de 
Schr\"odinger reside en su naturaleza de tipo parab\'olico, lo cual la hace 
complicada de resolver num\'ericamente. Esta complicaci\'on consiste en que se 
necesitan m\'etodos impl\'icitos para tener es\-ta\-bi\-li\-dad num\'erica. Por ello, proponemos 
hacer uso de la transformaci\'on de Madelung \cite{madelung:1926} para expresar 
nuestro sistema en una forma hidrodin\'amica, en la cual est\'an presentes un potencial 
de auto-interacci\'on y un potencial tipo cu\'antico\footnote{En la representaci\'on 
hidrodin\'amica de la ecuaci\'on de Schr\"odinger, el \'unico t\'ermino donde 
aparece la constante de Planck es en el potencial cu\'antico y por ello su nombre. 
Cabe mencionar que por la analog\'ia de las ecuaciones a lo largo del trabajo a 
dicho t\'ermino lo llamaremos potencial de tipo cu\'antico.} que es no lineal en
la densidad. Esta representaci\'on es de tipo 
hiperb\'olico, y es posible implementar m\'etodos num\'ericos expl\'icitos en forma de 
leyes de conservaci\'on, lo cual no es tan costoso computacionalmente como los 
impl\'icitos. Otra alternativa es usar la t\'ecnica de hidrodin\'amica de part\'iculas 
suavizadas (SPH por sus siglas en ingl\'es), que fue dise\~nada para resolver las 
ecuaciones de Navier-Stokes. Otra ventaja de esta representaci\'on, es que podemos 
obtener de manera sencilla un criterio en que el sistema Schr\"odinger-Poisson es estable.

Con estos objetivos en mente, dividimos este trabajo como sigue. 
En la secci\'on \ref{ch:modelo_EH} se deduce el sistema Schr\"odinger-Poisson a 
partir de un modelo de materia oscura escalar. En la secci\'on \ref{ch:modelo_hidro} 
expresamos dicho sistema en su formulaci\'on hidrodin\'amica usando la transformaci\'on 
de Madelung. En la secci\'on \ref{ch:jeans} hacemos el estudio del an\'alisis de 
inestabilidad de Jeans, el cual nos da un criterio para ver bajo qu\'e condiciones el 
sistema Schr\"odinger-Poisson es inestable. 
Despu\'es se dan las conclusiones donde se resumen las principales aportaciones de 
este trabajo. 
%

%%%%%%%%%%%%%%%%%%%%%%%%%%%%%%%%%%%%%%%%%%%%
%%%%%%%%%% SECTION MODELO_EH%%%%%%%%%%%%%%%%%%%%%%%%%
%\section[Modelo de Einstein-Hilbert]{El modelo de Einstein-Hilbert}\label{ch:modelo_EH}

\section[Campo d\'ebil]{Aproximaci\'on de campo d\'ebil de las ecuaciones de Einstein}\label{ch:modelo_EH}

La hip\'otesis que vamos a estudiar es que la materia oscura es un campo escalar complejo gobernado por la ecuaci\'on de Klein-Gordon

\begin{equation}\label{eq:kg}
\nabla_\mu \nabla^\mu \phi - \left( \frac{mc}{\hbar} \right)^2 \phi = 0\, ,
\end{equation}
y la respectiva ecuaci\'on para $\phi^*$, que se obtiene al calcular el complejo conjugado de (\ref{eq:kg}). Aqu\'{i} $\hbar$ es la constante de Planck, $c$ es la velocidad de la luz y $m$ es 
la masa del campo escalar.
Una manera de obtener la ecuaci\'on de Klein-Gordon la podemos 
ver en \cite{greiner_RQM:1990, Rider:1996}.

Dado que en relatividad general, al campo gravitacional se le asocia con la curvatura del 
espacio-tiempo, entonces campos gravitacionales d\'ebiles corresponder\'an a 
espacios-tiempos aproximadamente planos. En tal caso consideremos
\begin{equation}\label{eq:aprox_metrica}
g_{\mu \nu} = \eta_{\mu \nu} + h_{\mu \nu}
\end{equation}
con $|h_{\mu \nu}|\ll 1$ y $\eta_{\mu \nu}=\mbox{diag}(-1,1,1,1)$ la m\'etrica de Minkowski. 
Donde $g_{\mu \nu}$ es la m\'etrica 
del espacio-tiempo, que la descomponemos como una m\'etrica
plana m\'as una 
perturbaci\'on. 
Aqu\'i estamos en la libertad de poner una norma para $\bar{h}^{\mu \nu}$. Si pedimos que
\begin{equation}
\partial_\nu \partial^\nu \xi^\mu = \partial_\nu \bar{h}^{\nu \mu}
\end{equation}
lo cual es una ecuaci\'on de onda para $\xi^\mu$ con fuente $\partial_\nu \bar{h}^{\nu \mu}$, obtenemos que
\begin{equation}
\partial_\nu \bar{h}^{'\nu \mu} = 0\, .
\end{equation}
Tomando esta norma, las ecuaciones de Einstein quedan
\begin{equation}
\partial_\alpha \partial^\alpha \bar{h}_{\beta \nu} = - \kappa T_{\beta \nu} \,,
\end{equation}
donde $T_{\beta \nu}$ es el tensor de energ\'\i a-momento y
$\partial_\alpha \partial^\alpha$ es el operador D'Alambertiano en el espacio plano, 
es decir, es el operador 
que aparece en la ecuaci\'on 
de una onda. 
Estas son las ecuaciones de campo para un campo 
gravitacional d\'ebil en su forma est\'andar. A esta expresi\'on se le conoce 
como \textit{l\'imite de 
campo d\'ebil}.

Una aplicaci\'on particularmente importante de la aproximaci\'on de campo d\'ebil es el 
l\'imite newtoniano de la relatividad general. Este l\'imite no s\'olo corresponde a campos 
d\'ebiles sino tambi\'en a peque\~nas velocidades de las fuentes, 
en donde se satisface que $T^{00}\gg T^{ij}$\cite{Weinberg1972}.
En tal caso podemos considerar s\'olo la componente $\bar{h}^{00}$ e ignorar los dem\'as 
t\'erminos \cite{alcubierre:2008}. Para un an\'alisis detallado de la aproximaci\'on newtoniana 
de las ecuaciones de Einstein v\'ease \cite{argelia_tesis:2007}, la cual es la referencia de donde 
nos basaremos para presentar los siguientes resultados. Las ecuaciones de campo se 
reducen entonces a 
\begin{equation}
\partial_\alpha \partial^\alpha \bar{h}^{00} = -\frac{16\pi G}{c^2} m^2 \phi\phi^* \,,
\end{equation}
donde hemos tomado $\kappa =32\pi G/c^4$ y $T_{00}\approx \frac{1}{2}m^2c^2 \phi\phi^*$. 
Adem\'as, 
para velocidades peque\~nas, la derivada temporal, en el operador D'Alambertiano, 
es m\'as peque\~na que las derivadas espaciales,
as\'i que la ecuaci\'on se reduce a
\begin{equation} \label{eq:poisson_previa}
\nabla^2 \bar{h}^{00} = - \frac{16\pi G}{c^2} m^2 \phi\phi^* \,,
\end{equation}
donde el operador $\nabla^2$ es el Laplaciano 3-dimensional. Comparando este 
resultado con la ecuaci\'on de campo de Newton $\nabla^2 U = 4\pi G \rho$ concluimos que 
en este l\'imite $\bar{h}^{00}= -4U/c^2$ 
y $\bar{h}^{i0} = \bar{h}^{ij}=0$ (para $i \neq j$, $i,j=1,2,3$) y $m^2 \phi\phi^*$ 
corresponde a la fuente. Regresando a la definici\'on de $\bar{h}_{\alpha \beta}$ en 
t\'erminos de $h_{\alpha \beta}$, tenemos que $h^{00}= h^{ii}= -2U/c^2$. La expresi\'on 
final del l\'{i}mite new\-to\-nia\-no de las ecuaciones de Einstein la presentaremos 
en (\ref{eq:poisson}) debido que a\'un nos falta ver c\'omo se reduce $\phi$ en este 
l\'imite y lo mostraremos en la siguiente secci\'on. %\\

La m\'etrica del espacio-tiempo en la aproximaci\'on newtoniana es
\begin{equation}\label{eq:metrica_newton}
ds^2 = -(c^2+2U)dt^2 + (1-2U/c^2)d\vec{r}\cdot d\vec{r} \,,
\end{equation}
donde $d\vec{r}\cdot d\vec{r}=dx^2+dy^2+dz^2$. Cabe mencionar que el par\'ametro 
que tomaremos para hacer 
la
expasi\'on en serie es la cantidad adimensional $U/c^2\ll1$, lo cual es la aproximaci\'on 
a campo d\'ebil.

\subsection{La ecuaci\'on de tipo Schr\"odinger}

En esta secci\'on haremos la aproximaci\'on newtoniana de la ecuaci\'on de Klein-Gordon 
dando lugar a una ecuaci\'on tipo Schr\"odinger. Por simplicidad tomemos en los c\'alculos 
de esta secci\'on $c=\hbar=1$ y retomaremos las unidades originales del resultado bajo 
ciertos cambios que abajo daremos. Sustituyendo (\ref{eq:metrica_newton}) en (\ref{eq:kg})
\begin{eqnarray} \label{eq:aprox_kg}
g^{00}\partial_0\partial_0 \phi + g^{ii}\partial_i\partial_i \phi - m^2\phi &=&  \nonumber \\
-(1+2U)^{-1}\partial^2_t \phi + (1-2U)^{-1}\nabla^2 \phi -m^2\phi &\approx&  \nonumber \\
(1-2U)\partial^2_t \phi - (1 + 2U)\nabla^2 \phi + m^2\phi &=& 0 \,.
\end{eqnarray}

Para hacer de manera correcta el l\'imite newtoniano de la ecuaci\'on de 
Klein-Gordon tomemos\footnote{Una alternativa para realizar la aproximaci\'on 
newtoniana de la ecuaci\'on de Klein-Gordon se puede ver en la 
referencia \cite{argelia_tesis:2007}.} \cite{greiner_RQM:1990}
\begin{equation} \label{eq:cambio_variable1}
 \phi = \psi(\vec{x}, t) e^{-imt} \,.
\end{equation}

Sustituyendo en (\ref{eq:aprox_kg}) y teniendo en cuenta que $U\ll 1$, obtenemos 
\begin{equation}
-\frac{1}{2m}\ddot{\psi} + i\dot{\psi} + \frac{1}{2m} \nabla^2 \psi - mU \psi = 0
\end{equation}
donde $\ddot{\psi}=\partial^2_t\psi$ y $\dot{\psi}=\partial_t\psi$. 
Para un campo que var\'ia lentamente en el tiempo se tiene que $ \ddot{\psi}\approx 0 $, 
por lo que 
 \begin{equation} \label{eq:schrodinger}
i\hbar \dot{\psi} = -\frac{\hbar^2}{2m}\nabla^2 \psi + mU \psi\,,
\end{equation}
la cual es una ecuaci\'on que tiene la forma de la ecuaci\'on de Schr\"odinger y 
por ende la llamaremos \textit{ecuaci\'on de tipo Schr\"odinger}.
Adem\'as,
 hemos
regresado a las unidades originales haciendo $m\rightarrow mc^2/\hbar$, $\nabla^2\rightarrow c^2\nabla^2$ y $U\rightarrow U/c^2$.

La raz\'on del por qu\'e se dice que la ecuaci\'on (\ref{eq:schrodinger}) es de tipo 
Schr\"odinger y no la ecuaci\'on de Schr\"odinger que describe la mec\'anica cu\'antica 
es muy simple, la naturaleza del campo $\phi$ en (\ref{eq:kg}) es cl\'asico y por lo tanto 
tambi\'en lo es $\psi$. Sin embargo, los m\'etodos de soluci\'on que se usan para resolver 
la ecuaci\'on de Schr\"odinger pueden ser aplicados aqu\'{i}.

Usando (\ref{eq:cambio_variable1}) las ecuaciones de campo de 
Einstein (\ref{eq:poisson_previa}) se reducen a la \textit{ecuaci\'on de Poisson}
\begin{equation}\label{eq:poisson}
\nabla^2 U = 4\pi G \rho \,,
\end{equation}
siendo $U$ el potencial gravitacional debido al campo escalar con fuente 
$\rho=m^2 \psi\psi^*$. Al conjunto acoplado dado por (\ref{eq:schrodinger}) 
y (\ref{eq:poisson}) se le conoce en la literatura como \textit{sistema Schr\"odinger-Poisson}. 
Algunos estudios de este sistema los podemos ver 
en \cite{Guzman/Urena:2003, Bernal:2006it, Guzman/Urena:2004}.

%%%%%%%%%%%%%%%%%%%%%%%%%%%%%%%%%%%%%%%%%%%%
%%%%%%%%%% SECTION MODELO_HIDRO%%%%%%%%%%%%%%%%%%%%%%%
\section[Modelo hidrodin\'amico]{Modelo hidrodin\'amico del sistema Schr\"odinger-Poisson} \label{ch:modelo_hidro}
Tomando como motivaci\'on la formulaci\'on hidrodin\'amica de la ecuaci\'on de 
Sch\-r\"o\-din\-ger que rige la mec\'anica cu\'antica debida a Madelung \cite{madelung:1926}, posteriormente ampliada por 
Bohm \cite{bohm:1952}, expresaremos (\ref{eq:schrodinger}) en 
una representaci\'on de tipo hidrodin\'amico. En esta formulaci\'on de Madelung, 
la ecuaci\'on de Schr\"odinger, la cual es una ecuaci\'on diferencial lineal y compleja, 
se reemplaza por una densidad de probabilidad y su campo de velocidades.

\subsection{Versi\'on hidrodin\'amica de la ecuaci\'on \\de tipo Schr\"odinger}
Aplicando la transformaci\'on de Madelung \cite{madelung:1926}
\begin{equation} \label{eq:wave_function}
 \psi = R(\vec{r}, t) e^{ i S(\vec{r}, t) }\,,
\end{equation}
donde $R$ y $S$ son funciones reales, en 
(\ref{eq:schrodinger})
y despu\'es de algo de \'algebra se obtiene una ecuaci\'on compleja donde la parte 
ima\-gi\-na\-ria est\'a dada por
\begin{equation}\label{part_imag}
\hbar \partial_t R = -2\frac{\hbar^2}{2m} \nabla S\cdot \nabla R -\frac{\hbar^2}{2m} R \nabla^2 S\,,
\end{equation}
y la correspondiente parte real es
\begin{equation}\label{eq:part_real}
-\hbar R \partial_t S = mU R - \frac{\hbar^2}{2m} \nabla^2 R + \frac{\hbar^2}{2m} R(\nabla S)^2\,.
\end{equation}

Es conveniente definir las siguientes variables de campo
\begin{eqnarray}
\vec{v} &=& \frac{\hbar}{m}\nabla S \label{eq:velocity}\,, \\
\rho &=& m^2R^2 \label{eq:density}\,.
\end{eqnarray}

Ahora, multiplicando (\ref{part_imag}) por $R$ y usando la 
identidad $2R\nabla S \cdot \nabla R = \nabla \cdot(R^2 \nabla S) - R^2 \nabla^2 S$ y 
las definiciones (\ref{eq:velocity}) y (\ref{eq:density}) obtenemos la ecuaci\'on de continuidad
\begin{equation}\label{eq:continuity}
\partial_t \rho + \nabla \cdot(\rho \vec{v}) = 0 \,,
\end{equation}
siendo $\rho$ una densidad y $\rho \vec{v}$ un flujo de la densidad $\rho$ que entra o 
sale de una regi\'on. Ahora dividiendo por $R$ y luego aplicando 
el gradiente a la ecuaci\'on (\ref{eq:part_real}) y usando la 
identidad $\nabla(\nabla S \cdot \nabla S) = 2(\nabla S \cdot \nabla)\nabla S$ y 
las definiciones (\ref{eq:velocity}) y (\ref{eq:density}) obtenemos
\begin{equation}\label{eq:euler}
\partial_t \vec{v} + (\vec{v}\cdot \nabla)\vec{v} = -\nabla \left( U - \frac{\hbar^2}{2m^2} \frac{\nabla^2 \sqrt{\rho}}{\sqrt{\rho}}\right )\,.
\end{equation}

Las ecuaciones (\ref{eq:continuity}) y (\ref{eq:euler}) son la \textit{formulaci\'on 
hidrodin\'amica} de la ecuaci\'on de campo de tipo Schr\"odinger. Estas ecuaciones 
indican que la evoluci\'on temporal del campo $\psi(\vec{r}, t)$ es equivalente al flujo 
de un ``fluido'' de densidad $\rho(\vec{r}, t)$ cuyas part\'iculas de masa $m$, se mueven 
con una velocidad $\vec{v}(\vec{r}, t)$ sujetos a una fuerza derivada de un potencial 
externo $U(\vec{r}, t)$ m\'as una fuerza adicional debido al potencial 
cu\'antico\footnote{En analog\'ia a la representaci\'on hidrodin\'amica de la 
ecuaci\'on de Schr\"odinger, el potencial $Q$ lo llamaremos potencial cu\'antico.} 
$Q(\vec{r}, t)$, el cual depende de la densidad del 
fluido \cite{madelung:1926, weiner_askar(54)8_1971, weiner_askar(54)3_1971} 
y lo definimos como
\begin{equation}
Q(\vec{r}, t) \equiv - \frac{\hbar^2}{2m^2}\frac{\nabla^2 \sqrt{\rho}}{\sqrt{\rho}}\,.
\end{equation}

Es justo el potencial cu\'antico el que hace la diferencia entre un sistema hidrodin\'amico cl\'asico y el sistema cu\'antico que estamos estudiando. La ecuaci\'on (\ref{eq:euler}) es justo la de Burgers si despreciamos este pontencial tomando el l\'{i}mite cl\'asico $\hbar\rightarrow 0$.
Observemos que $S$ es la fase de la funci\'on de onda $\psi$, la cual es univaluada, es decir, de (\ref{eq:wave_function}) es inmediato ver que $S$ es una funci\'on peri\'odica tal que en todo punto $S(p') = S(p) + 2\pi n$, y por lo tanto,
\begin{equation}\label{eq:compatibility}
\oint m \vec{v}\cdot d\vec{r} = \hbar \oint \nabla S \cdot d\vec{r} = \hbar \int_p^{p'} dS = 2 \pi \hbar n \,, 
\end{equation}
con $n= 0, \pm 1, \pm 2,\dots $ para toda trajectoria ``cerrada''. Esto implica que el sistema hidrodin\'amico tiene v\'ortices, algo que ya es conocido y observado en condensados de Bose-Einstein en el laboratorio (ver por ejemplo \cite{pethick}). Este resultado es una \textit{condici\'on de 
compatibilidad} entre la versi\'on hidrodin\'amica (\ref{eq:continuity}) 
y (\ref{eq:euler}) y la ecuaci\'on de tipo Schr\"odinger, esto es, 
una soluci\'on $(\rho, \vec{v})$ le corresponde una $\psi$ bien definida y de 
manera un\'ivoca.

No obstante que el fluido descrito por el sistema (\ref{eq:euler}) es irrotacional debido a que la velocidad es un gradiente, el ``fluido'' descrito por (\ref{eq:continuity}) y (\ref{eq:euler}) 
tiene una diferencia esencial respecto a un fluido ordinario: si en la ecuaci\'on
 (\ref{eq:compatibility}) hacemos la aproximaci\'on de que a lo largo de la trayectoria
 de integraci\'on usamos la magnitud promedio de $\vec{v}$  a lo largo de la trayectoria,
 tenemos que
 en un movimiento 
rotacional, la magnitud de $\vec{v}$ decrece cuando la distancia al centro crece, y viceversa. 
Esto es debido a la condici\'on de compatibilidad (\ref{eq:compatibility}) (ver \cite{pethick}). 
M\'as detalles y discusi\'on sobre estos puntos pueden ser consultados en \cite{Pitaevskii2003,Wilhelm1970}.

\subsection[Hidrodin\'amica: Schr\"odinger-Poisson]{Hidrodin\'amica: Schr\"odinger-Poisson}

La din\'amica del campo escalar se convierte 
en un sistema hidrodin\'amico, constituido por ecuaciones semejantes 
a las que describen a un fluido no viscoso. El conjunto Schr\"odinger-Poisson 
en esta formulaci\'on es
\begin{eqnarray}
\partial_t \rho + \nabla \cdot(\rho \vec{v}) &=& 0 \label{eq:SSchP_continuidad} \\
\partial_t \vec{v} + (\vec{v}\cdot \nabla)\vec{v} &=& -\nabla \left( U - \frac{\hbar^2}{2m^2} \frac{\nabla^2 \sqrt{\rho}}{\sqrt{\rho}}\right ) \label{eq:SSchP_euler}\\ 
\nabla^2 U &=& 4\pi G \rho \label{eq:SSchP_poisson} \,.
\end{eqnarray}

Este sistema representa la din\'amica de un ``fluido'' auto interactuante en equilibrio con 
un potencial tipo cu\'antico. Haciendo una analog\'ia estructural de las ecuaciones 
nos damos cuenta que no tenemos un t\'ermino de presi\'on como en los fluidos normales, 
sino que tenemos un t\'ermino que depende de la densidad,
por lo que es razonable 
pensar que estamos tratando con una ``ecuaci\'on de estado'' barotr\'opica. En la 
literatura cuando la presi\'on $p=0$ se dice que el sistema es polvo, y aunque no es muy 
realista, es buena aproximaci\'on en algunos casos especiales. As\'\i \ es como se modela
la materia oscura en el modelo $\Lambda$CDM.

Sin embargo, dado que la fuerza de tipo cu\'antico es opuesta a la fuerza 
de auto-interacci\'on, podemos pensar que es debido a una presi\'on efectiva del fluido 
y expresarla como 
\begin{equation}
-\nabla Q = \frac{1}{\rho} \nabla p_q(\rho)=\frac{\hbar^2}{4m^2\rho}\nabla({\nabla^2 \ln{\rho}}) \,,
\end{equation}
donde $p_q \neq 0$ es una presi\'on efectiva.

%%%%%%%%%%%%%%%%%%%%%%%%%%%%%%%%%%%%%%%%%%%%
%%%%%%%%%% SECTION JEANS%%%%%%%%%%%%%%%%%%%%%%%%%
\section{Ecuaciones hidrodin\'amicas en el regimen lineal e inestabilidad de Jeans}\label{ch:jeans}

En esta secci\'on estudiamos la estabilidad del sistema 
Sch\-r\"o\-din\-ger-Poisson en el r\'egimen lineal en su forma hidrodin\'amica. Este an\'alisis est\'a 
basado en el estudio de la inestabilidad de Jeans para el colapso gravitacional de un 
gas protoestelar (m\'as detalles pueden ser consultados en \cite{Chavanis2011}).

James Jeans demostr\'o que en un fluido homog\'eneo e isotr\'opico se pueden 
generar  
peque\~nas fluctuaciones en la densidad y en la velocidad \cite{jeans:1929}. 
Sus c\'alculos fueron hechos en el contexto de un 
fluido est\'atico. En particular mostr\'o que las fluctuaciones de densidad 
pueden crecer en el tiempo 
si los gradientes de presi\'on que ocurren en esa regi\'on donde est\'a la fluctuaci\'on
de densidad son menores que a la auto-interacci\'on gravitacional.
No es sorprendente que tal efecto exista ya que la gravedad 
es una fuerza atractiva, si las fuerzas de presi\'on son m\'as peque\~nas que las
gravitatorias, una regi\'on con mayor densidad que la de fondo atrae 
materia de sus alrededores y llega a ser m\'as densa. Mientras m\'as 
densa es la regi\'on, m\'as materia atraer\'a, resultando en el colapso de una 
fluctuaci\'on de la densidad y generando un objeto ligado gravitacionalmente. 
El criterio para ver cuando puede ocurrir este colapso o no es calculando la 
longitud de Jeans del fluido y lo explicamos enseguida.

Antes de calcular la longitud de Jeans 
con el suficiente detalle matem\'atico,
hagamos un c\'alculo de orden de magnitud para ver su significado f\'isico. 
Consideremos que en un instante dado hay una fluctuaci\'on esf\'erica de 
radio $\lambda$, homog\'enea, densidad positiva $\rho_1>0$ y masa $M$ 
situado en un medio de densidad media $\rho_0$. Las fluctuaciones 
crecer\'an si la fuerza de auto-interacci\'on por unidad de masa, $F_g$, excede 
la fuerza opuesta por unidad de masa que surge de la presi\'on $F_p$
\begin{equation}
F_g \approx \frac{GM}{\lambda^2} \approx {G\rho \lambda}> F_p \approx \frac{p}{\rho \lambda} \approx \frac{v_s^2}{\lambda}\,,
\end{equation}
donde $v_s$ es la velocidad del sonido. Esta relaci\'on implica que la fluctuaci\'on 
crece si $\lambda > v_s(G\rho)^{-1/2}$. Esto establece la existencia de la longitud 
de Jeans  $\lambda_J \approx v_s(G\rho)^{-1/2}$.

\subsection{Ecuaciones hidrodin\'amicas linealizadas}

Para hacer el estudio de la inestabilidad de Jeans es necesario linealizar el sistema 
hidrodin\'amico 
(\ref{eq:SSchP_continuidad})--(\ref{eq:SSchP_poisson}).
Este sistema de ecuaciones admite la soluci\'on est\'atica con 
$\rho=\rho_0$, $\vec{v}=0$ y $\nabla U=0$. Sin embargo, si $\rho \neq 0$, 
el potencial gravitacional deber\'ia variar espacialmente, es decir, una 
distribuci\'on homog\'enea de $\rho$ 
est\'atica puede ser inestable ante perturbaciones y puede
estar globalmente expandi\'endose o contray\'endose\cite{Weinberg1972}. La incompatibilidad de un 
Universo est\'atico con el principio cosmol\'ogico en gravedad 
newtoniana se da tambi\'en en el universo 
est\'atico de Einstein. De cualquier modo, cuando consideramos un 
universo en expansi\'on los resultados de Jeans no cambian cualitativamente. 
Para ver esto busquemos una soluci\'on al sistema 
hidrodin\'amico 
(\ref{eq:SSchP_continuidad})--(\ref{eq:SSchP_poisson}) 
que representa una peque\~na perturbaci\'on a la soluci\'on est\'atica  
\begin{equation}\label{eq:perturbaciones}
\rho \approx \rho_0 + \epsilon \rho_1\,, \hspace{1cm} \vec{v} \approx \epsilon \vec{v}_1\,, \hspace{1cm} U \approx U_0 + \epsilon U_1\,,
\end{equation}
donde las variables de perturbaci\'on $\rho_1$, $\vec{v}_1$ y $U_1$ son funciones 
del espacio y tiempo y $\epsilon \ll 1$. Sustituyendo (\ref{eq:perturbaciones}) en 
(\ref{eq:SSchP_continuidad})--(\ref{eq:SSchP_poisson}) 
y tomando los t\'erminos a primer orden de $\epsilon$ obtenemos el respectivo 
conjunto linealizado
\begin{eqnarray}
\partial_t \rho_1 + \rho_0 \nabla \cdot \vec{v}_1 &=& 0 \label{eq:continuity_lineal} \,,\\
\partial_t \vec{v}_1 &=& -\nabla U_1 + \frac{\hbar^2}{4m^2}\nabla \left[ \nabla^2 (\rho_1/\rho_0)  \right] \label{eq:euler_lineal}\,, \\
\nabla^2 U_1 &=& 4\pi G \rho_1 \label{eq:poisson_lineal} \,.
\end{eqnarray}

Estudiemos las soluciones del sistema linealizado (\ref{eq:continuity_lineal})-(\ref{eq:poisson_lineal}) buscando soluciones en la forma de ondas planas
\begin{equation} \label{eq:delta_ui}
\delta u_i = \delta_{i0}e^{i(\vec{k}\cdot \vec{r}+\omega t)} \,,
\end{equation}
donde las perturbaciones $\delta u_i$ con $i=1,2,3$ corresponden a 
$\rho_1$, $\vec{v}_1$ y $U_1$, y las 
amplitudes 
$\delta_{i0}$ 
las denotaremos por
$D$ para la densidad, $\vec{V}$ para la velocidad y por $\mathcal{U}$  
para el potencial gravitacional.
El vector $\vec{r}$ es el vector posici\'on, $\vec{k}$ es el vector de onda 
y $\omega$ es la frecuencia angular de oscilaci\'on, el cual en general 
es 
un n\'umero
complejo. Usando (\ref{eq:delta_ui}) en (\ref{eq:continuity_lineal})-(\ref{eq:poisson_lineal}) 
y definiendo $\delta_0 \equiv D/\rho_0$ obtenemos
\begin{eqnarray}
\omega \delta_0 + \vec{k}\cdot \vec{V} &=& 0 \, \label{eq:pertubada_continuidad} \\
\omega \vec{V} + \mathcal{U}\vec{k} + \frac{\hbar^2}{4m^2}\delta_0 k^2 \vec{k} &=& 0 \,, \label{eq:perturbada_euler}\\
k^2 \mathcal{U} + 4\pi G \rho_0 \delta_0 &=& 0 \,. \label{eq:perturbada_poisson}
\end{eqnarray}

Brevemente consideremos las soluciones con $\omega=0$, es decir, las que no 
dependen del tiempo. Es evidente de (\ref{eq:pertubada_continuidad}), que el 
vector de onda $\vec{k}$ es perpendicular a la velocidad $\vec{v}_1$. 
Juntando estas expresiones obtenemos
\begin{equation}
k = \left( \frac{16\pi G \rho_0 m^2}{\hbar^2} \right)^{1/4}\,.
\end{equation}

Ahora tratemos las soluciones dependientes del tiempo, $\omega \neq 0$. 
Derivemos (\ref{eq:continuity_lineal}) res\-pec\-to a $t$ y considerando que 
$\vec{v}_1$ es una funci\'on tal que 
$ \partial_t \nabla \cdot \vec{v}_1 = \nabla \cdot (\partial_t \vec{v}_1)$, y luego sustituimos 
las expresiones (\ref{eq:euler_lineal}) y (\ref{eq:poisson_lineal}) para obtener
\begin{equation}\label{eq:onda_rho}
\partial_t^2 \rho_1 -  4\pi G\rho_0\rho_1 + \frac{\hbar^2}{4m^2}\nabla^2 \nabla^2 \rho_1 = 0\,.
\end{equation}
Sustituyendo (\ref{eq:delta_ui}) en (\ref{eq:onda_rho}) se llega a una ``relaci\'on de dispersi\'on''
\begin{equation}
\omega^2 = \frac{\hbar^2}{4m^2}k^4 - 4\pi G \rho_0 \,.
\end{equation}
Es importante se\~nalar que la ecuaci\'on (\ref{eq:onda_rho}) no es una ecuaci\'on de onda.

Esta relaci\'on de dispersi\'on es an\'aloga a la que encontr\'o Jeans \cite{jeans:1929,Kolb:1990}, 
siendo la diferencia que la forma del potencial cu\'antico hace que la magnitud del vector 
de onda tenga exponente $4$. Para soluciones estables, pedimos que $\omega>0$, lo 
cual implica 
\begin{equation} 
 k > k_J \equiv \left( \frac{16\pi G \rho_0 m^2}{\hbar^2} \right)^{1/4} \,,
\end{equation}
donde definimos el vector de onda de Jeans $k_J$. En base a este resultado la longitud 
de onda de Jeans, $\lambda_J = 2\pi/k_J$, est\'a dada por
\begin{equation}\label{eq:Jeans_longitud}
\lambda_J = \left(\frac{\pi^3\hbar^2}{G\rho_0 m^2}\right)^{1/4}\,.
\end{equation}

Entonces para longitudes de onda mayores que la longitud de onda de Jeans tenemos 
que la perturbaci\'on crece exponencialmente. En base a la definici\'on de 
la longitud de Jeans, reescribimos la relaci\'on de dispersi\'on como
\begin{equation}
\omega = \pm \frac{\hbar}{2m}k^2 \left[ 1 - \left(\frac{\lambda}{\lambda_J}\right)^4 \right]^{1/2}\,.
\end{equation}
Recordemos que la velocidad del sonido en la teor\'ia de Jeans se define 
como $v_s^2\equiv \partial p/\partial\rho$. Por analog\'{i}a en nuestro caso definimos una cantidad 
que llamaremos la velocidad cu\'antica del sonido o velocidad de grupo \footnote{Esta cantidad tiene unidades de velocidad, pero no es en realidad una velocidad, es s\'olo una manera de llamarla. La velocidad de grupo se define como $v_g\equiv \partial \omega / \partial k$ y nos 
describe la velocidad de una perturbaci\'on f\'\i sica. 
Si $\lambda \ll \lambda_J$ entonces $\tilde{v}_s$ si coincide con la expresi\'on formal de velocidad 
de grupo. Debemos se\~nalar que $\tilde{v}_s$ al tener unidades de velocidad es una 
velocidad de referencia o t\'\i pica en el sistema y su correspondencia con velocidad de sonido o de grupo debe tomarse con reserva.} 
como
\begin{equation}\label{eq:vel_sonido}
\tilde{v}_s \equiv \frac{\hbar}{m}k \,.
\end{equation}
De (\ref{eq:delta_ui}), (\ref{eq:pertubada_continuidad})-(\ref{eq:perturbada_poisson}) 
y (\ref{eq:vel_sonido}) obtenemos
\begin{eqnarray}
\frac{\rho_1}{\rho_0} &=& \delta_0 e^{i(\vec{k}\cdot \vec{r} \pm |\omega| t)} \,,\\
\vec{v}_1 &=& \mp \frac{1}{2} \frac{\vec{k}}{k} \tilde{v}_s \delta_0 \left[ 1-\left( \frac{\lambda}{\lambda_J} \right)^4 \right]^{1/2} e^{i(\vec{k}\cdot \vec{r} \pm |\omega| t)} \,, \\
U_1 &=& - \delta_0 \frac{1}{4} \tilde{v}_s^2 \left( \frac{\lambda}{\lambda_J} \right)^4 e^{i(\vec{k}\cdot \vec{r} \pm |\omega| t)} \,.
\end{eqnarray}
Cuando $\lambda>\lambda_J$ la frecuencia $\omega$ es imaginaria
\begin{equation}
\omega = \pm i (4\pi G \rho_0)^{1/2}\left[ 1 - \left( \frac{\lambda_J}{\lambda} \right)^4 \right]^{1/2}\,.
\end{equation}
En este caso las expresiones anteriores quedan
\begin{eqnarray}
\frac{\rho_1}{\rho_0} &=& \delta_0 e^{i\vec{k}\cdot \vec{r} \pm |\omega| t} \,,\\
\vec{v}_1 &=& \mp i\frac{\vec{k}}{k^2} \delta_0 (4\pi G\rho_0)^{1/2} \left[ 1-\left( \frac{\lambda_J}{\lambda} \right)^4 \right]^{1/2}
\nonumber \\ & & \times
 e^{i\vec{k}\cdot \vec{r} \pm |\omega| t} \,, \\
U_1 &=& - \frac{1}{4} \delta_0 \tilde{v}_s^2 \left( \frac{\lambda}{\lambda_J} \right)^4 e^{i\vec{k}\cdot \vec{r} \pm |\omega| t} \,.
\end{eqnarray}
las cuales representan una soluci\'on de ondas estacionarias, cuyas amplitudes 
crecen en el tiempo para el caso $e^{i\vec{k}\cdot \vec{r} + |\omega| t}$. 
La escala de tiempo caracter\'istico 
para la 
evoluci\'on de esta amplitud se define como
\begin{equation}
\tau \equiv |\omega|^{-1} = (4\pi G \rho_0)^{-1/2}\left[ 1 - \left( \frac{\lambda_J}{\lambda} \right)^4 \right]^{-1/2} \,.
\end{equation}
Para escalas $\lambda \gg \lambda_J$, el tiempo caracter\'istico $\tau$ coincide con 
el tiempo de colapso de ca\'ida libre\footnote{El tiempo de colapso libre es el tiempo 
necesario para que un sistema colapse bajo su propia atracci\'on gravitacional en 
ausencia de fuerzas opuestas.}, 
dado por
$\tau_{ff} = (4\pi G\rho_0)^{-1/2}$, pero cuando $\lambda \longrightarrow \lambda_J$, 
este tiempo diverge.

Se define la masa de Jeans como la masa contenida en una esfera de 
radio $\lambda_J/2$ y densidad media $\rho_0$, esto es
\begin{eqnarray}
M_J = \frac{1}{6}\pi\rho_0 \left( \frac{\pi^3\hbar^2}{G\rho_0 m^2} \right)^{3/4} \,.
\end{eqnarray}
La masa de Jeans es la masa m\'inima de la perturbaci\'on para que \'esta aumente 
con el tiempo. 

Un estudio de la inestabilidad de Jeans tanto en el contexto cosmol\'ogico como el local 
referente al modelo de axiones para materia oscura se encuentra en \cite{Sikivie:2009}. En este trabajo se
menciona que en el r\'egimen lineal, los condensados de Bose-Einstein de axiones y 
CDM  son indistinguibles para todas las escalas de inter\'es observacional. Sin embargo 
en el r\'egimen no lineal de formaci\'on de estructura, estas dos teor\'ias difieren en el 
potencial cu\'antico.

\subsection{Estimaci\'on de la masa del campo escalar}

Despejando la masa del campo escalar de la ecuaci\'on (\ref{eq:Jeans_longitud}) obtenemos
\begin{equation}
m = \left( \frac{\pi^3 \hbar^2}{G\rho_0 \lambda_J^4} \right)^{1/2} \,,
\end{equation}
Hagamos un c\'alculo de orden de magnitud para la masa del campo escalar, vamos a utilizar
\begin{eqnarray}
\hbar &\sim& 10^{-34}\, \mbox{J}\cdot \mbox{s} \rightarrow \hbar^2 \sim 10^{-68}\, \mbox{J}^2 \cdot \mbox{s}^2 \,, \\
G &\sim& 10^{-11} \, \mbox{m}^3/\mbox{Kg}\cdot \mbox{s}^2 
\end{eqnarray}

Seg\'un la teor\'ia de formaci\'on de estructura est\'andar, un valor de la longitud de 
Jeans $\lambda_J$ puede ser fijado para la \'epoca en que la radiaci\'on y la materia 
son iguales teniendo un factor de escala $a\simeq 1/3200$. En este tiempo, 
$\rho \simeq 3 \times 10^{10} \mbox{M}_\odot a^{-3}/\mbox{Mpc}^3$ \cite{wmap:2007}, 
y la longitud de Jeans $\sim \mathcal{O}(10^2)\mbox{pc}$ \cite{Silverman:2002}. 
Aqu\'i $\mbox{M}_\odot =1.9891\times 10^{30} \mbox{Kg}$ es la masa del Sol. 
Entonces tomando la densidad $\rho_0 \sim 10^{-14}\mbox{Kg}/\mbox{m}^3$ 
y $\lambda_J\sim 10^{2}\mbox{pc}$ tenemos
\begin{eqnarray}
m   &\sim& 10^{-57} \mbox{Kg}
\end{eqnarray}
o en unidades de $e\mbox{V}/\mbox{c}^2$ obtenemos
\begin{equation}
m \sim 
10^{-22}e\mbox{V}/\mbox{c}^2 \,.
\end{equation}
Por lo tanto, con el c\'alculo de la estabilidad de Jeans, necesitamos que la masa del 
campo escalar sea $\sim10^{-22}e\mbox{V}/\mbox{c}^2$, lo cual es un resultado obtenido 
en \cite{Matos:2002, Lee/Lim:2010}. De la definici\'on de la masa de Jeans obtenemos
\begin{equation}
 M_J \sim 10^{10} \mbox{M}_\odot \,.
\end{equation}

%%%%%%%%%%%%%%%%%%%%%%%%%%%%%%%%%%%%%%%%%%%%
%%%%%%%%%% SECTION CONCLUSION%%%%%%%%%%%%%%%%%%%%%%%%
\section{Conclusiones}

En este trabajo  
hemos revisado
el modelo de materia oscura escalar para des\-cri\-bir halos de materia oscura. 
A partir de la acci\'on de Einstein-Hilbert acoplado con un campo escalar complejo se hizo la aproximaci\'on de campo d\'ebil y luego se calcul\'o el l\'imite para campos que var\'ian 
lentamente en el tiempo conocido como l\'imite newtoniano. El sistema resultante se 
conoce como \textit{sistema Schr\"odinger-Poisson}. Luego, mediante la transformaci\'on 
de Madelung, la ecuaci\'on de Schr\"odinger, ecuaci\'on lineal y compleja, se convierte en 
un sistema no lineal que describe la din\'amica de campos reales, una densidad y su campo 
de velocidades: \textit{formulaci\'on hidrodin\'amica}. Cabe mencionar que dicha transformaci\'on 
no altera la ecuaci\'on de Poisson.

En el contexto hidrodin\'amico, aparece de manera clara un potencial de tipo cu\'antico 
el cual genera una fuerza opuesta a la fuerza gravitacional. Entonces mediante un estudio 
de la inestabilidad de Jeans se obtuvo un criterio para que una perturbaci\'on de la densidad
 colapse y estas fuerzas se equilibren. Bajo esta condici\'on y valores adecuados se calcul\'o 
 la masa del campo escalar.

Por otro lado, la din\'amica de las variables de campo es de naturaleza hiperb\'olica lo cual 
es de suma importancia para la implementaci\'on de esquemas en forma de leyes de 
conservaci\'on usando m\'etodo expl\'icitos de diferencias finitas, los cuales se piensan usar en futuros trabajos. Adem\'as, dada la 
similitud con las ecuaciones de Navier-Stokes, podemos usar las t\'ecnicas que han 
sido dise\~nadas para resolverlas como el m\'etodo de hidrodin\'amica de part\'iculas 
suavizadas (SPH). Este \'ultimo m\'etodo es muy interesante porque no requiere la 
implementaci\'on de condiciones a la frontera (num\'ericas). Este trabajo est\'a en curso y 
ser\'a reportado en un futuro.

El sistema Schr\"odinger-Poisson con simetr\'ia esf\'erica, se ha estudiado en el contexto de 
estrellas de bosones y condensados de Bose-Einstein, sin embargo no en la versi\'on 
hidrodin\'amica. 
Las soluciones 
num\'ericas con dicha simetr\'\i a se abordar\'an en trabajos futuros dado que su soluci\'on representa un trabajo 
complejo. 

%%%%%%%%%%%%%%%%%%%%%%%%%%%%%%%%%%%%%%%%%%%%
%%%%%%%%%% SECTION BIBLIOGRAPHY%%%%%%%%%%%%%%%%%%%%%

\end{document}